\documentclass[%
reprint,
 amsmath,amssymb,
 aps,
]{revtex4-2}

\usepackage{graphicx}
\usepackage{dcolumn}
\usepackage{bm}
\usepackage{placeins}

\usepackage{xspace}
\usepackage{color}
\newcommand{\tss}[1]{\ensuremath{^{\text{#1}}}}
\newcommand{\rrscan}{r\tss{2}SCAN\xspace}
\renewcommand{\v}[1]{\ensuremath{\mathbf{#1}}}

\begin{document}


\title{Reliable lattice dynamics from an efficient density functional}

\author{Jinliang Ning}
\email{jning1@tulane.edu}
\affiliation{Department of Physics and Engineering Physics, Tulane University, New Orleans, Louisiana 70118, United States}

\author{James W. Furness}
\affiliation{Department of Physics and Engineering Physics, Tulane University, New Orleans, Louisiana 70118, United States}

\author{Jianwei Sun}
\email{jsun@tulane.edu}
\affiliation{Department of Physics and Engineering Physics, Tulane University, New Orleans, Louisiana 70118, United States}

\date{\today}

\begin{abstract}

First principles predictions of lattice dynamics are of vital importance for a broad range of topics in materials science and condensed matter physics. The large-scale nature of lattice dynamics calculations and the desire to design novel materials with distinct properties demands that first principles predictions are accurate, transferable, efficient, and reliable for a wide variety of materials. In this work, we demonstrate that the recently constructed \rrscan density functional meets this need for general systems by demonstrating phonon dispersions for typical systems with distinct chemical characteristics.  The functional's performance opens a door for phonon-mediated materials discovery from first principles calculations. 



\end{abstract}

\maketitle


\section{\label{sec:Intro}Introduction}

Each new age of human technology has been enabled by the discovery of new materials. From the knapped stone and simple metallurgy of history to the semiconductor revolution of recent decades, a new understanding of materials has expanded the horizons of possibility. This trend for materials to drive progress has not gone unnoticed and ever increasing effort is being devoted to using computational models to search the vast materials space for desirable new compounds.

Phonons are the quanta of lattice waves driven by the elementary thermal excitation of the atoms or molecules that make up a condensed matter system. Intuitively, long-wavelength phonons are perceived as sound. Phonons can interact with electronic structure and have a profound impact on a wide range of observed material phenomena, from thermal and electrical conductivity through to more exotic charge density waves and superconductivity, alongside their decisive role controlling the dynamic stability of materials. This position at the center of materials property design has driven prediction of phonon spectra to become an important aspect of materials space searches.

The connection between the vibrational frequency of a phonon, $\omega(k)$, and the wavevector, $k$, is known as the phonon dispersion. It can be measured experimentally by inelastic neutron or x-ray scattering. The phonon dispersion can also be predicted from theory using force constants calculated with computational models, though the cost of such calculations is generally high. This results in a simultaneous requirement for phonon calculations to efficiently scale for high-throughput workflows while maintaining sufficient accuracy to usefully guide experiments. Density functional theory (DFT) \cite{KohnShamDFT} using a semi-local exchange-correlation (XC) functional offers an appealing balance of these considerations and has become the workhorse computational method for high throughput materials discovery. Efficient evaluation of phonon spectra can be obtained from density functional calculations using density functional perturbation theory \cite{baroni2001phonons}, or through direct displacements of the atoms \cite{DirectFC}. Indeed, density functional methods have already proved effective tools for identifying new phonon phenomena, with the discovery of high/room temperature hydrogen-based superconductors as a prominent example\cite{H2S_prediction,H3S_expt}. \\
\indent High throughput calculations of thermodynamic properties \cite{MaterialsProject} is vital for phase diagram predictions \cite{CALPHAD,thermocalc}, and discovery of new meta-stable materials \cite{ning_MBT} places a particularly high demand for simultaneous accuracy, transferability, efficiency, and reliability. While the accuracy of a DFT calculation is largely determined by the accuracy of the chosen XC functional, the high computational demand of phonon calculations largely excludes expensive nonlocal XC functionals, like hybrid density functionals \cite{HSE06}. The current choice for phonon calculations remains conventional density functionals, including local density approximation (LDA) and the Perdew-Burke-Ernzerhof (PBE) generalized gradient approximation (GGA). While efficient and reasonably accuracy, one problem with these conventional density functionals is their transferability in the materials space where different compounds can have very different chemical bonds. Recent progress has shown that semi-local meta-GGAs can maintain this efficiency while being accurate for a wide variety of materials \cite{SCAN,SCAN_NChem, remsing2017Si, yubostripe,James_SCAN_Cuprates, yubo_MO,Yubo_TiO2,Peng_MnO2,Peng_MO,ning_MBT, Zhang2017a, lane2018antiferromagnetic, zhang2020SmB6}, exemplified by the strongly constrained and appropriately normed (SCAN) meta-GGA \cite{SCAN,SCAN_NChem}. Unfortunately, extensive use has shown that SCAN suffers numerical problems that are exaggerated in phonon calculations, making reliably obtaining accurate phonon spectra from SCAN calculations a challenging task. Here, we show that a revised version of SCAN which solves the numerical problems, called \rrscan \cite{Furness2020c}, delivers accurate, transferable, and reliable lattice dynamics. This is demonstrated in a selected set of materials that have different bonding characteristics. We then further explain the origins of such excellent performance of r2SCAN for phonon calculations.

\begin{figure*}
\centering
\includegraphics[width=\linewidth]{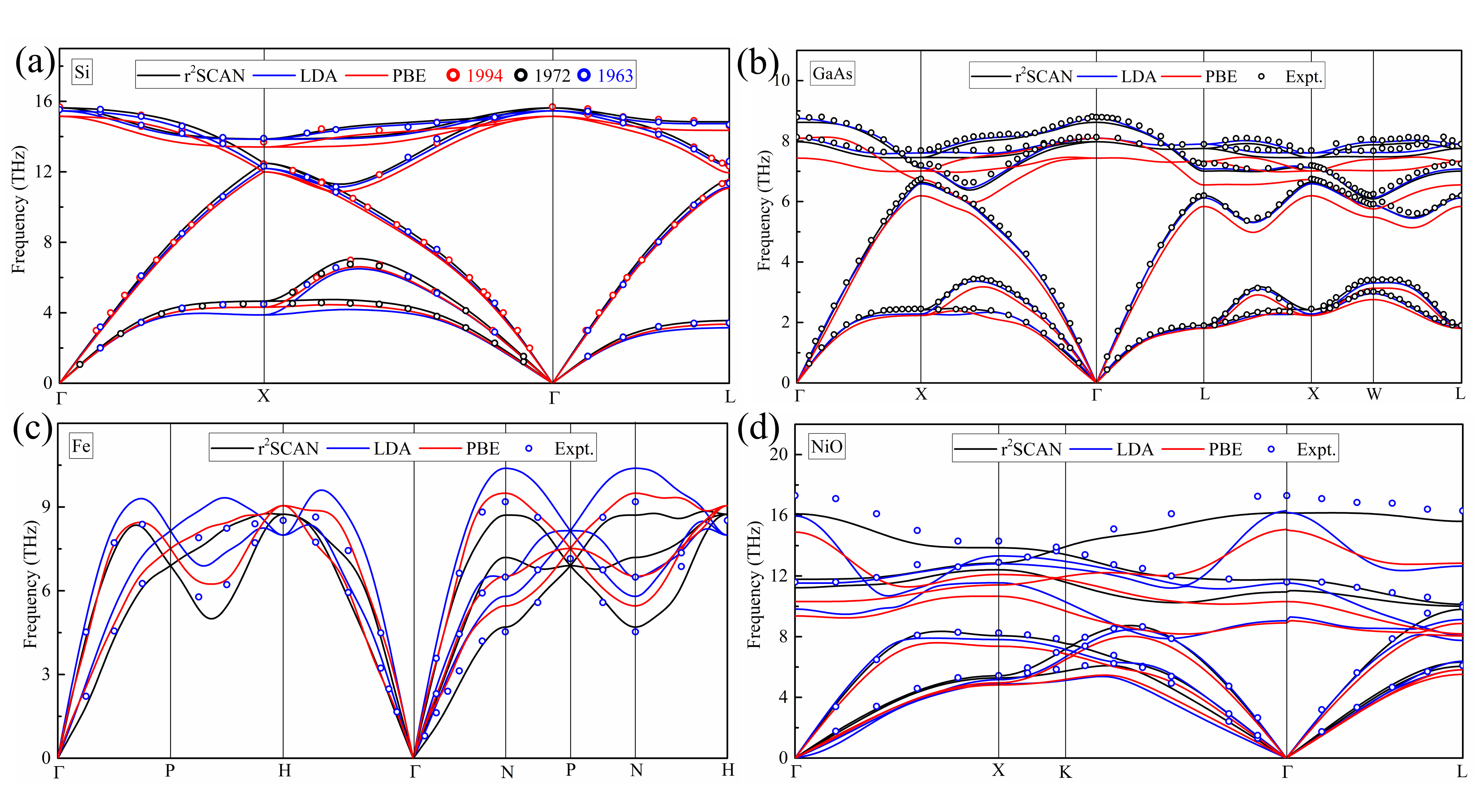}
\caption{Well converged Phonon dispersions of (a) Si, (b) GaAs, (c) Fe, and (d) NiO, calculated by LDA, PBE, and r2SCAN, compared with available experimental data, of 1963 \cite{Si1963}, 1972 \cite{Si1972}, and 1994 \cite{Si1994} for Si, of 1990 \cite{GaAs_expt} for GaAs, from Ref. \cite{Fe_expt} for Fe, and from Ref. \cite{NiO_expt} for NiO. The second 3 experimental acoustic band data points along $\Gamma$-K direction for NiO were directly taken from their figure of Ref. \cite{NiO_expt}. \label{fig:ideal}}
\end{figure*}

\section{\label{sec:Results}Results}
The selected small test set includes four solids. Two are industrially important semiconductors with covalent or mixed covalent-ionic interactions: covalent Si in the diamond structure, and GaAs with zinc blende structure. One magnetic metal: body-centered-cubic (bcc) Fe. We also include a magnetic oxide with covalent-ionic bonding NiO. These solids have been widely studied experimentally and theoretically and their phonon dispersions have been accurately determined from experiments. In all cases we are interested in establishing both the accuracy and numerical stability of the phonon spectra calculated by different density functionals.

\subsection{\label{subsec: Si and GaAs}Si and GaAs}
Figures \ref{fig:ideal} a) and b) compare calculated phonon dispersions with experimental results for Si and GaAs respectively. The aim of these calculations is to establish the relative accuracy of the functionals under ideal conditions with high-accuracy computational settings tuned to ensure well converged results for all functionals. We find that LDA predicts a relatively accurate spectrum for GaAs but underestimates lower frequency phonon bands in Si, while PBE underestimates phonon frequencies across the board. The \rrscan meta-GGA shows the most consistent accuracy across both materials and at all energy ranges, closely matching the experimental data.

\begin{figure}
\centering
\includegraphics[width=\linewidth]{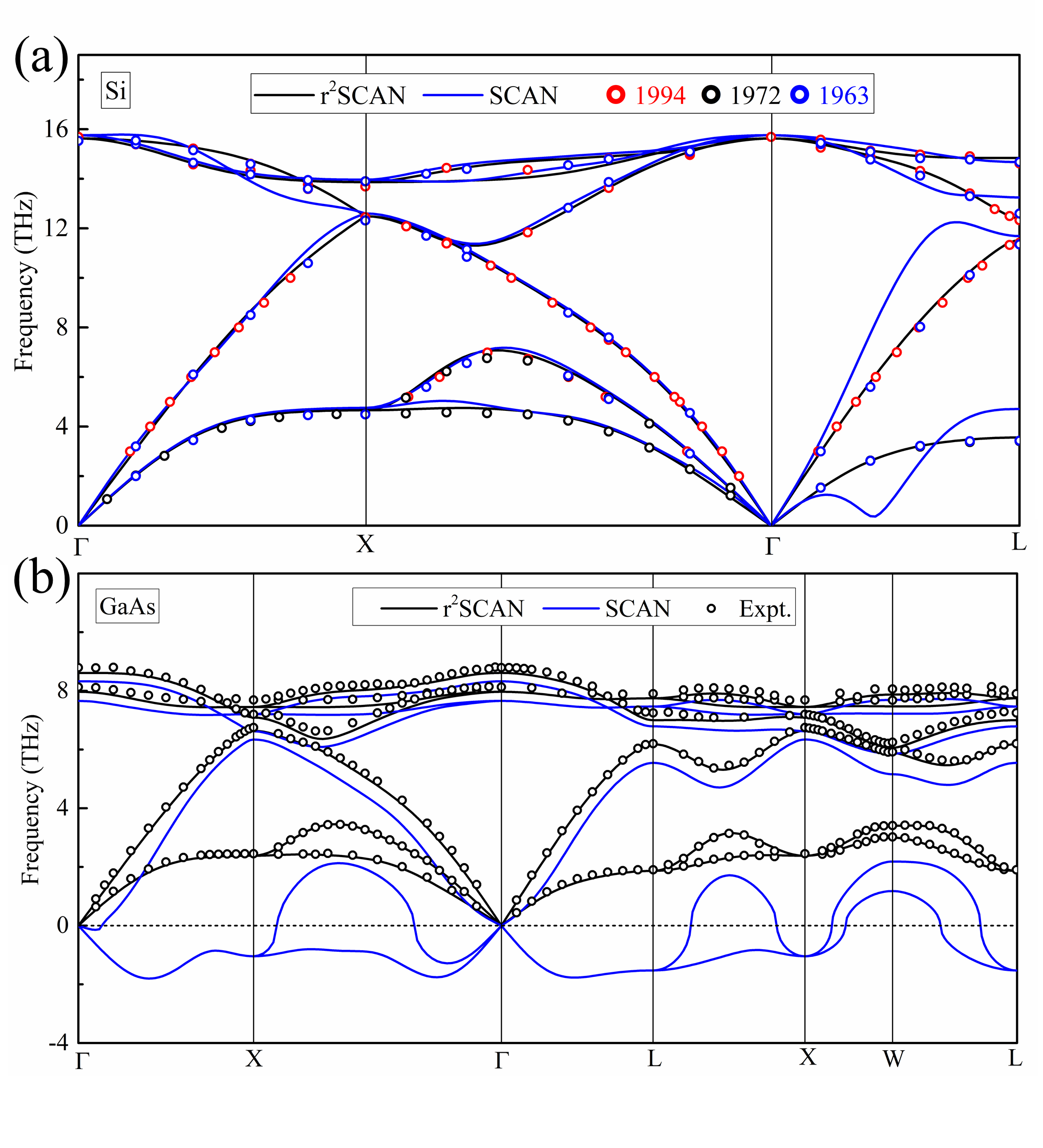}
\caption{Phonon dispersions of (a) Si and (b) GaAs calculated by SCAN and r2SCAN with default settings, compared with available experimental data from 1963 \cite{Si1963}, 1972 \cite{Si1972}, and 1994 \cite{Si1994} for Si, and from 1990 \cite{GaAs_expt} for GaAs. \label{fig:default}}
\end{figure}

To establish the relative efficiency of \rrscan compared to its parent SCAN functional, we repeat the calculations of Figures \ref{fig:ideal} a) and b) using the default VASP computational settings that better reflect a high-throughput workflow. The resulting phonon spectra are shown in Figures \ref{fig:default} a) and b) for Si and GaAs respectively. For Si of Figure \ref{fig:default} a), SCAN and \rrscan show similar accuracy across much of the spectrum, though SCAN's error is significant for low frequency bands between the $L$ and $\Gamma$ points. For GaAs in Figure \ref{fig:default} b) however, the numerical problems of the SCAN functional are immediately apparent. Here spurious imaginary frequencies occur across the SCAN spectrum and the higher frequency bands show generally poor accuracy. Conversely, the \rrscan functional remains well behaved under these cheaper settings, predicting an accurate and well converged spectrum for high and low frequencies. Note that imaginary frequencies are predicted by SCAN despite using a fully relaxed ionic structure that should be stable along all wave vectors. This prediction of spurious imaginary frequencies can be attributed to incomplete sampling of sharp oscillations in the SCAN XC potential as the ionic positions are displaced \cite{Furness2020c, Price2021}. With special tuning of the parameters like Fourier transform grid density and atomic displacement size, SCAN can deliver accurate phonon dispersions for these two solids, as shown in the supplementary materials. Such tuning tricks just highlight the serious numerical problems of SCAN due to potential surface oscillations however. It is not guaranteed that these tuning tricks can solve SCAN's numerical problems for all solids.

\begin{table*}
\caption{Lattice constants and transition metal local magnetic moment $m$ for Si, GaAs, Fe, Ni and NiO, calculated from different functionals and compared with experimental data. $\epsilon_{\infty}$ and Z* are the high-frequency dielectric constants and the diagonals of Born effective charge (the values in parentheses are for the z direction component), used for non-analytical term corrections for phonon dispersions of GaAs and NiO. $\Delta$ZPE is the correction to lattice constants due to the zero point energy.}
\label{tab:lattice}
\begin{ruledtabular}
\begin{tabular}{llllllllllll}
	&	Si	&	\multicolumn{3}{c}{GaAs}	&	\multicolumn{2}{c}{Fe}		&	\multicolumn{5}{c}{NiO}	\\
\cline{2-2}
\cline{3-5}
\cline{6-7}
\cline{8-12}
Methods	&	a	&	a	&	Z*	&	$\epsilon_{\infty}$ 	&	a	&	m	&	a	&	c	&	Z*	&	$\epsilon_{\infty}$	&	m	\\
\cline{1-12}
LDA	&	5.4029	&	5.6110	&	2.07	&	17.33	&	2.7470	&	1.93	&	2.8828	&6.9706	&2.09(2.74)	&	44.9(46.0)	&	1.08	\\
PBE	&	5.4688	&	5.7505	& 2.32		& 92.83 &	2.8304	&	2.18	&	2.9687	&	7.2260	&	2.24 (2.67)	&	21.2(21.98)	&	1.35	\\
SCAN	&	5.4273	&	5.6670	&	2.16	& 12.67	&	2.8424	&2.61	&	2.9445	&	7.1874	&	2.18(2.26)	&	6.81 (6.90)	&	1.58	\\
\rrscan	&	5.4398	&	5.6688	&	2.15	&	11.63	&	2.8629	&	2.71&2.9461	&	7.1884	&	2.18(2.29)	&	7.53(7.63)	&	1.56\\
Expt.	&	5.4309$^a$	&	5.6556$^b$	&	2.18$^j$	& 10.89$^k$		&	2.8608$^c$	&	2.13$^d$	&	2.9517$^e$	&	7.2170$^e$	&	2.2$^g$	&	5.7$^h$	&	1.90$^f$	\\
Expt. - $\Delta$ZPE$^i$	&	5.422	&	5.641	&		&		&	2.855	&		&		&		&		&		&		\\

\end{tabular}
\end{ruledtabular}
\raggedright 
$a$\cite{Si_LC} $b$\cite{GaAs_LC} $c$\cite{Fe_LC} 79 K 	$d$\cite{FeNi_m} 	$e$\cite{Ni_LC} 78 K	$e$\cite{NiO_LC} 10 K $f$\cite{NiO_m} $g$\cite{NiO_born} $h$\cite{NiO_epsilon} $i$\cite{ZPE_LC} $j$\cite{GaAs_Z} $k$\cite{GaAs_epsilon} 
\end{table*}

\subsection{\label{subsec:Fe}Fe}
Figure \ref{fig:ideal} c) compares calculated and experimental phonon dispersions for bcc Fe. Here, the improvement from GGA to meta-GGA is less clear, with both showing regions of accuracy. The $N$ high-symmetry point of bcc Fe appears particularly challenging for all functionals. It is notable however, that \rrscan significantly improves over other functionals for the lower band along $\Gamma-N-P-N-H$, though this trend is reversed at higher frequencies with PBE showing greater accuracy for the middle and higher bands.

It has commonly been observed that SCAN tends to overestimate magnetic moments and the magnetisation energy of simple magnetic metals like Fe, Co, and Ni \cite{Isaacs2018a, Jana2018, Romero2018, Ekholm2018, Fu2018, Fu2019, Mejia-Rodriguez2019c}. As we include bcc Fe we have calculated the local magnetic moments for the transition metal atoms and present the results in Table \ref{tab:lattice}. We see that \rrscan maintains the over-magnetisation of SCAN, as should be expected from \rrscan's construction as a regularisation of SCAN. Figure \ref{fig:ideal} c) shows that the \rrscan calculated phonon dispersions are not unduly degraded by this over-magnetisation however.

\subsection{\label{subsec:NiO}NiO}

Figure \ref{fig:ideal} d), compares the calculated and experimental phonon dispersions for NiO. The \rrscan meta-GGA shows significant improvements over LDA and PBE for this material. In particular, the high-frequency optical bands from LDA and PBE are qualitatively wrong, while \rrscan is reasonably accurate. Note we allow the crystal structure to be fully relaxed from the ideal AFM FCC structure, resulting in a small shrinkage of the lattice in the direction perpendicular to the ferromagnetic Ni planes. This symmetry breaking then leads to three optical bands. For this system we expect the main source of error to be self-interaction error, which causes the $d$ orbital electron density to become too diffuse and fractionally occupied. In comparison with LDA and PBE, SCAN and \rrscan reduce self-interaction errors and localize the $d$ electrons around the Ni ion to a greater degree \cite{yubo_MO,Yubo_TiO2}, stabilizing the magnetic moment as shown Table \ref{tab:lattice}. The self-interaction error of semi-local functionals can be remedied by including a Hubbard $U$ term. The \textit{ad hoc} nature of its parameterization limits predictive power however.

Polar bonds, such as those found in systems like NiO and GaAs, can cause the longitudinal optical and transverse optical (LO-TO) splitting in the experimentally observed phonon dispersion. A non-analytical correction to the phonon dispersion \cite{NAC,NAC_gonze,NAC2_gonze} based on the high-frequency dielectric constants $\epsilon_{\infty}$ and Born effective charge Z* must therefore be considered. Table \ref{tab:lattice} shows that \rrscan delivers slightly better Z* in comparison with LDA and PBE. As $\epsilon_{\infty}$ is related to the response of electrons to the external electric field and can be strongly affected by the self-interaction error, \rrscan significantly improves $\epsilon_{\infty}$ over LDA and PBE, although the discrepancy in $\epsilon_{\infty}$ from the experimental values is notable. In order to better illustrate the comparison of calculated force constants, we use the \rrscan $\epsilon_{\infty}$ and Z* for the non-analytical term corrections \cite{NAC,NAC_gonze,NAC2_gonze} of the phonon dispersions of NiO and GaAs for all functionals.

\section{Discussion}

Viewing Figure \ref{fig:ideal} as a whole, we can see a broad trend of accuracy across the different materials: PBE $<$ LDA $<$ \rrscan. It is perhaps surprising that the inclusion of gradient information into the PBE GGA results in worse accuracy than the simpler LDA functional for the phonon dispersions. This effect can also be viewed from the other direction: it is surprising that LDA is as successful as it is for the phonon dispersion, particularly given its well known tendency to underestimate bond lengths and lattice constants. The reason for this can be found in the force constant dynamic matrix \cite{baroni2001phonons},
\begin{align}
    \frac{\partial^2E(\v{R})}{\partial \v{R}_I \partial \v{R}_J} \equiv 
    &- \frac{\partial \v{F}_I}{\partial \v{R}_J} =
    \int \frac{\partial n_\v{R}(\v r)}{\partial \v{R}_J}\frac{\partial V_\v{R}(\v r)}{\partial \v R_I} d\v r \nonumber \\
    &+ \int n_\v R(\v r)\frac{\partial^2 V_\v R(\v r)}{\partial \v R_I\partial \v R_J}d \v r + \frac{\partial^2 E_N(\v R)}{\partial \v R_I \partial \v R_J},\label{eq:af_matrix}
\end{align}
where $\v R_I$ is the position of nucleus $I$, $n_\v R(\v r)$ and $V_\v R(\v r)$ are the ground state electron density and nuclear potential respectively with nuclei in positions $\v R$, and $E_N(\v R)$ is the Coulomb repulsion between the nuclei at positions $\v R$.

As the LDA bond lengths are too short, the second order derivative of the nuclear repulsion energy is overestimated (the final term of Eq. \ref{eq:af_matrix}). This error is compensated however, by an overestimation of the linear response of electron density $n_\v R(\v r)$ to the nuclear distortion in the first term of Eq. \ref{eq:af_matrix}. Since the first and final terms of Eq. \ref{eq:af_matrix}) have opposite signs, their errors are favorably cancelled. The overestimation of the linear response of electron density is a consequence of the self-interaction errors intrinsic to semilocal density functionals, including LDA, PBE, and \rrscan. PBE tends to overestimate bond lengths without correcting the linear response, so the favorable cancellation is lost. Like its parent functional \rrscan improves both these aspects, giving accurate lattice constants \cite{Furness2020c} while simultaneously improving linear response characteristics \cite{SCAN_NChem} as demonstrated in Table \ref{tab:lattice} for lattice constants and $\epsilon_{\infty}$. This results in a more accurate phonon spectrum with greater transferability across different classes of materials.

As previously mentioned, Figure \ref{fig:default} shows how \rrscan improves on the SCAN functional by avoiding the numerical sensitivities that necessitate the expensive tuning of the fast Fourier transform grid. A full analysis of the origin of the numerical issues in SCAN, and their solution in \rrscan, is presented in Ref. \cite{Furness2020c}. When calculating lattice dynamics from finite atomic displacements the smooth exchange-correlation potential of \rrscan is well sampled by a coarse grid while the sharp oscillations of the SCAN potential are not \cite{Furness2020c, Price2021}. This poor sampling results in slow and unpredictable convergence of the SCAN phonon spectrum with grid density, and the appearance of spurious imaginary frequencies.

\section{\label{sec:Conclusions}Conclusions}
We have tested the performance of \rrscan for calculating the phonon dispersions of typical systems relative to experimental data and other commonly used functionals (LDA and PBE). Our results for these systems suggest that \rrscan can calculate accurate lattice dynamics for general systems with good transferability between different bonding characteristics. Across all the materials tested we find \rrscan is either the best choice, or competitive with the best choice in the case of magnetic metals. While we find that SCAN can be accurate when Fourier transform grid and atomic displacement settings are tuned, however its poor numerical stability makes identifying the ideal parameters burdensome. Additionally, the necessary use of expensive Fourier transform grids prevents the SCAN functional being truly useful to high throughput studies. When default low-cost computational settings are used we find that SCAN predicts spurious imaginary bands. These problems are avoided in the \rrscan functional which predicts accurate phonon spectra even from low-cost default parameters. While we find that while LDA and PBE can be quite accurate for some systems, they do not show the same generally transferable accuracy as \rrscan does. With these inspiring findings, we strongly recommend \rrscan to the community as an effective computational tool for future phonon dispersion studies.

\section{\label{sec:Methods}Methods}
DFT \cite{KohnShamDFT} calculations with the LSDA, PBE\cite{Perdew1996}, SCAN\cite{SCAN}, and \rrscan\cite{Furness2020c} XC functionals were performed using the Vienna Ab-initio Simulation Package (VASP) \cite{VASP}. The projector-augmented wave (PAW) method was used to treat the core ion-electron interaction \cite{PAW,PAW_vasp}. An energy cutoff of 600 eV was used to truncate the plane wave basis. A $\Gamma$-centered mesh with a spacing threshold of 0.15 \AA$^{-1}$ was used for $k$-space sampling for unit cell relaxations of semiconducting systems Si, GaAs and NiO, and 0.1 \AA$^{-1}$ for metallic Fe. For supercell atomic force calculations, only a single $\Gamma$ point is used for semiconducting systems Si, GaAs and NiO, and a $2\times2\times2$ k-point mesh for metallic Fe. A Gaussian smearing with 0.02 eV is used for semiconducting systems Si, GaAs and NiO, and Methfessel--Paxton smearing with 0.2 eV for Fe. For atomic force calculations, $3\times3\times3$, $3\times3\times3$, $5\times5\times5$, and $4\times4\times2$ supercells of the conventional unit cells (as shown in Table 1) are used for Si, GaAs, Fe and NiO, respectively. The ionic positions of all systems were relaxed for all functionals until the maximum ionic forces were below 1 meV\AA$^{-1}$. We used the Phonopy code \cite{phonopy} to obtain the harmonic force constants from VASP atomic force calculations within finite displacement method (0.015 \AA). 
For Figure \ref{fig:ideal}, PREC = High; ENAUG = 2000 is specified for \rrscan, while for Figure \ref{fig:default}, we used the VASP officially recommended accurate defaults (PREC = Accurate) together with special tuned fast Fourier transform grid density for comparison. The full set of comparison is referred to the supplementary material. 


\begin{acknowledgments}

 J.N. and J.S. acknowledge the support of the U.S. DOE, Office of Science, Basic Energy Sciences Grant No. DE-SC0014208 and J.W.F. acknowledges the support of DE-SC0019350.

\end{acknowledgments}

\bibliography{apsnjl, phonon_paper}

\end{document}


\preprint{APS/123-QED}

\title{Supplementary: Reliable lattice dynamics from an efficient density functional}

\author{Jinliang Ning}
\email{jning1@tulane.edu}
\affiliation{Department of Physics and Engineering Physics, Tulane University, New Orleans, Louisiana 70118, United States}

\author{James W. Furness}
\affiliation{Department of Physics and Engineering Physics, Tulane University, New Orleans, Louisiana 70118, United States}

\author{Jianwei Sun}
\email{jsun@tulane.edu}
\affiliation{Department of Physics and Engineering Physics, Tulane University, New Orleans, Louisiana 70118, United States}

\date{\today}

\maketitle

Table 1 and Figures 1-2 contain the detailed information for the comparison of numerical stability between SCAN and \rrscan with respect to specific input settings, including the basic accuracy setting (the PREC tag in INCAR), the FFT-grid and the size of small displacements used in atomic force calculations, being tested on Si and GaAs. Figure 3 shows the phonon dispersions of GaAs and NiO calculated by LDA, PBE and \rrscan with the non-analytical correction using Born effective charge and high-frequency dielectric constant from the underlying density functional. In contrast, the Figures 1(b) and (d) in the main text use the \rrscan Born effective charge and high-frequency dielectric constants for all LDA, PBE and \rrscan results.

\begin{table*}
\caption{Parameter settings for VASP calculations of Si and GaAs phonon results by \rrscan and SCAN for Figure \ref{fig:sm_si} and \ref{fig:sm_gaas} in supplemental materials, respectively. The Tag column defines the tags used in plots in  the Figures. The PREC column gives the PREC tag value in INCAR. NGX and NGXF columns in the Basic settings specify how NGX and NGXF are determined by VASP accordingly. ``Specified'' means the NGX for the Supercell is manually specified in INCAR to be consistent with times of the expansion from the unit cell to supercells. The Unit cell and Supercell columns give the exact value of NGX and NGXF in turn. In addition, $d$ (in \AA) is the finite displacements of atoms in supercells used for atomic force calculations. Figure 2 of the main text uses the ``AccNGX'' settings. NGX, NGY and NGZ set the number of grid points in the FFT-grid along the three lattice vectors, while NGXF, NGYF, NGZF set the number of grid points in the ``fine'' FFT-grid  along the three lattice vectors. Due to the cubic symmetry of GaAs and Si, NGX = NGY = NGZ and NGXF = NGYF = NGZF.}
\label{tab:params}
\begin{ruledtabular}
\begin{tabular}{lllllllll}
 \multicolumn{9}{c}{Si} \\
\cline{1-9}
 &	\multicolumn{3}{c}{Basic Settings}		&	\multicolumn{2}{c}{Unit cell}		& \multicolumn{3}{c}{Supercell} \\
\cline{2-4}
\cline{5-6}
\cline{7-9}
Tag	&	PREC	&	NGX	&	NGXF	&	NGX	&	NGXF	&	NGX	&	NGXF	&	$d$	\\
\cline{1-9}
Acc	    &	Accurate	&	ENCUT	    &	2NGX	&	48	&	96	&	140	&	280	&	0.015	\\
AccNGX	&	Accurate	&	Specified	&	2NGX	&	48	&	96	&	144	&	288	&	0.015	\\
HighNGX\_d03	&	High	& Specified &	ENAUG	&	48	&	108	&	144	&	320	&	0.03	\\
\hline
\hline
\multicolumn{9}{c}{GaAs} \\
\cline{1-9}
 &	\multicolumn{3}{c}{Basic Settings}		&	\multicolumn{2}{c}{Unit cell}		& \multicolumn{3}{c}{Supercell} \\
\cline{2-4}
\cline{5-6}
\cline{7-9}
Tag	&	PREC	&	NGX	&	NGXF	&	NGX	&	NGXF	&	NGX	&	NGXF	&	$d$	\\
\cline{1-9}
Acc	&	Accurate	&	ENCUT	&	2NGX	&	48	&	96	&	140	&	280	&	0.015	\\
AccNGX	&	Accurate	&	Specified	&	2NGX	&	48	&	96	&	144	&	288	&	0.015	\\
High	&	High	&	ENCUT	&	ENAUG	&	48	&	112	&	140	&	336	&	0.015	\\
High\_d03	&	High	&	ENCUT	&	ENAUG	&	48	&	112	&	140	&	336	&	0.03	\\
\end{tabular}
\end{ruledtabular}
\end{table*}

\begin{figure}
\includegraphics[width=\linewidth]{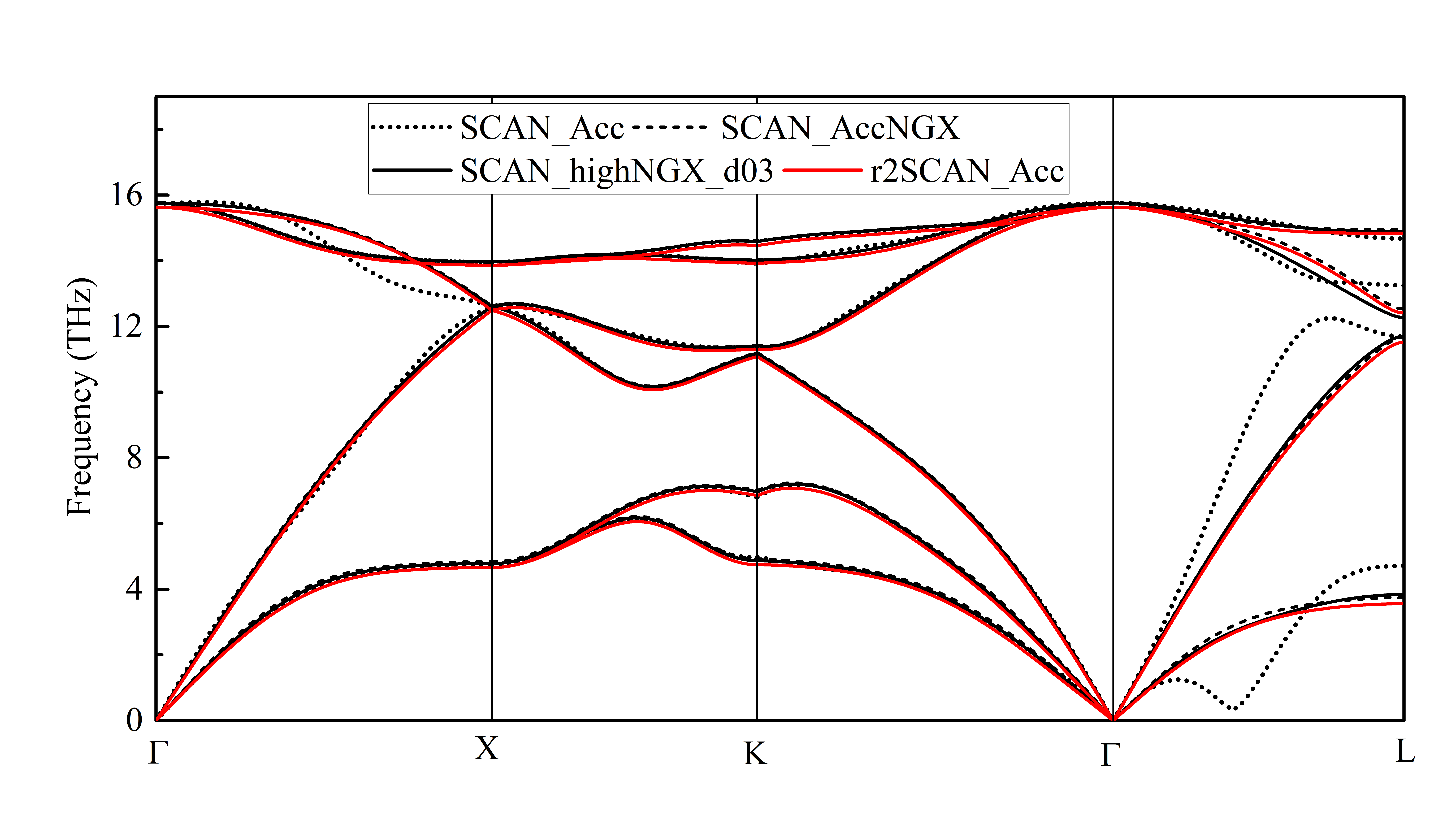}
\caption{Phonon dispersions of Si calculated by SCAN and \rrscan. See Table \ref{tab:params} for definition of legend tags. \label{fig:sm_si}}
\end{figure}

\begin{figure}
\centering
\includegraphics[width=\linewidth]{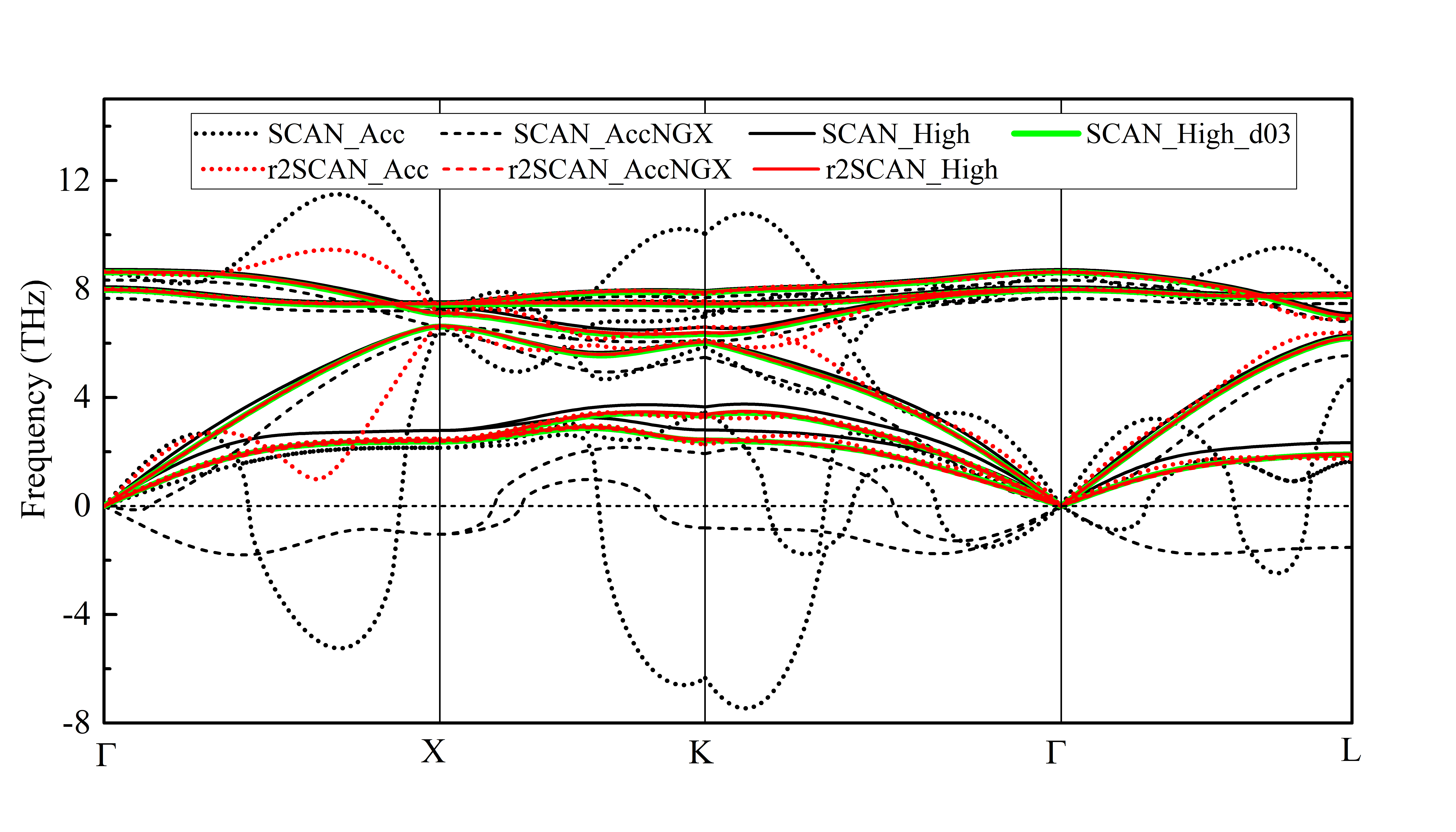}
\caption{Phonon dispersions of GaAs calculated by SCAN and \rrscan. See Table \ref{tab:params} for definition of legend tags. \label{fig:sm_gaas}}
\end{figure}

\begin{figure}
\centering
\includegraphics[width=\linewidth]{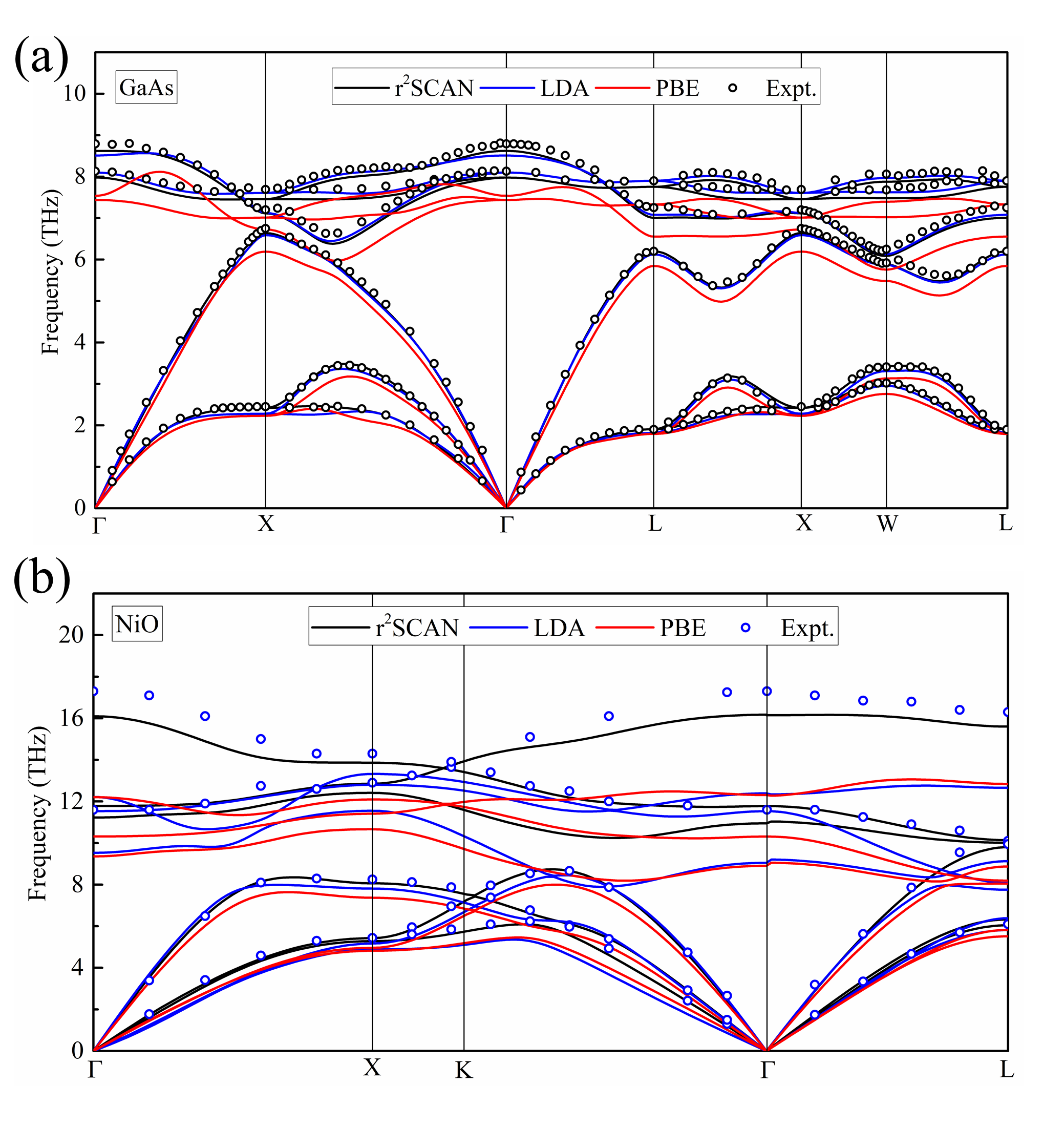}
\caption{Phonon dispersions of (a) GaAs and (b) NiO of LDA, PBE and \rrscan with the non-analytical correction using Born effective charge and high-frequency dielectric constants calculated by the underlying density functional, compared with experimental results. \label{fig:sm_gaas_NiO}}
\end{figure}
